\documentstyle[12pt,a4,epsfig,a4wide]{article} 

\def\Jo#1#2#3#4{{#1} {\bf #2}, #3 (#4)}

\def\NPB{{Nucl. Phys.} {\bf B}}

\def\PLB{{Phys. Lett.}  {\bf B}}
\def\PRL{Phys. Rev. Lett.}
\def\PRD{{Phys. Rev.} {\bf D}}

\def\EPC{{Eur. Phys. J.} {\bf C}}

\def\IJMP{Int. J. Mod. Phys. {\bf A}}

\def\JHEP{JHEP}
\def\ra{\rightarrow}
\def\be{\begin{equation}}
\def\ee{\end{equation}}
\def\gs{\mathrel{
   \rlap{\raise 0.511ex \hbox{$>$}}{\lower 0.511ex \hbox{$\sim$}}}}
\def\ls{\mathrel{
   \rlap{\raise 0.511ex \hbox{$<$}}{\lower 0.511ex \hbox{$\sim$}}}}
\newcommand{\obb}{0\mbox{$\nu\beta\beta$}}
\newcommand{\onbb}{neutrinoless double beta decay}

\newcommand{\ba}{\begin{array}{c}}
\newcommand{\baz}{\begin{array}{cc}}
\newcommand{\bad}{\begin{array}{ccc}}
\newcommand{\bav}{\begin{array}{cccc}}
\newcommand{\bea}{\begin{equation} \begin{array}{c}}
\newcommand{\eea}{ \end{array} \end{equation}}
\newcommand{\ea}{\end{array}}
\newcommand{\D}{\displaystyle}

\newcommand{\meff}{\mbox{$\langle m \rangle$ }}

\newcommand{\dma}{\mbox{$\Delta m^2_A$}}
\newcommand{\dms}{\mbox{$\Delta m^2_\odot$}}

\begin{document}

\title{  
\hfill { \bf {\small hep-ph/0207053}}\\ 
\hfill { \bf {\small DO--TH 02/09}}\\ \vskip 1cm 
\bf Leptogenesis, mass hierarchies and low energy parameters}
\author{Werner Rodejohann\thanks{E--mail: \tt 
rodejoha@xena.physik.uni-dortmund.de}\\ \\
{\normalsize \it Institut f\"ur Theoretische Physik, 
Universit\"at Dortmund,}\\
{\normalsize \it Otto--Hahn--Str.4, 44221 Dortmund, Germany}}
\date{}
\maketitle
\thispagestyle{empty}
\begin{abstract}
Leptogenesis in a left--right symmetric model is investigated for 
all possible neutrino mass hierarchies. 
The predictions of the model 
for low energy parameters as measured 
in neutrinoless double beta decay and in oscillation experiments are 
compared. The preferred values of the Majorana phases and limits 
on the smallest mass state are given.  
The main observation is 
that for the inverse hierarchy observable $CP$ violation in oscillation 
experiments as well as a sizable signal in 
neutrinoless double beta decay can be expected. 
In a degenerate scheme, one Majorana phase is bounded to be around 
$\pi/2$ or $\pi$, and this ambiguity can easily be tested through 
neutrinoless double beta decay. 
The dependence of the baryon asymmetry 
on the different ``Dirac'' and ``Majorana'' phases is analyzed and a 
possibility to avoid the gravitino problem is discussed.  
\end{abstract}

\newpage

\section{\label{sec:intro}Introduction}
The connection of leptogenesis \cite{leptogenesis} with low energy 
parameters has been investigated in a number of 
recent publications \cite{others,hiki,branco2,ellis}. 
Typically, a see--saw mechanism 
\cite{seesaw} connects 
the light neutrinos as indicated by oscillation experiments with 
the heavy Majorana neutrinos whose decay creates the observed baryon 
asymmetry of the universe. 
In \cite{JPR1} this very connection was analyzed within 
left--right symmetric 
models, for which a very simple connection between 
leptogenesis and neutrino oscillations was found 
\cite{JPR2}\footnote{See, e.g., \cite{zhang} 
for the possibility of 
baryogenesis in left--right symmetric models.}. 
This simple picture yields identical low and high energy sectors of the 
theory. 
Predictions of the scenario for \meff$\!\!$, the effective Majorana mass 
of the electron neutrino, were given. Its predicted value lies around 
$10^{-3}$ eV, as expected in a normal hierarchical neutrino mass scheme. 
The allowed values of the $CP$ violating phases in the mixing matrix were 
investigated in \cite{apu}. In the present note we generalize these 
previous works by including also the inverse and degenerate 
mass hierarchies. 
In addition, we use a more appropriate fit \cite{hiki} for the solution of 
the Boltzmann equations.  
Apart from the fact that in the inverse hierarchy 
\meff is now considerably larger than in the 
normal hierarchical scheme, the predictions for other low energy 
parameters differ, be it the preferred value of the leptonic 
Jarlskog parameter or the lower limit on the smallest 
mass eigenstate. The 
identical low and high energy sectors of the theory 
allow to show the resonance 
effect for degenerate neutrinos in a very simple manner. 
A two--fold ambiguity of the value of one Majorana phase 
is observed, which can easily be resolved through \onbb. Finally, the 
gravitino problem is commented on and a possibility to avoid it in 
our scenario is discussed. 

The paper is organized as follows: In Section \ref{sec:oscleplr} we 
shortly review the formalism of leptogenesis in left--right symmetric models.  
We then estimate the baryon asymmetry in Section \ref{sec:estYB} in 
both the normal and inverse hierarchy as well as for degenerate neutrinos.  
The full numerical results are given in Section \ref{sec:res} and we 
finally conclude in Section \ref{sec:concl}.

\section{\label{sec:oscleplr}The model and leptogenesis}
The gauge group of left--right symmetric models is  
$SU(2)_L \times SU(2)_R \times U(1)_{B - L}$ and the leptonic 
mass term reads: 
\be
\mbox{${\cal L}$} = 
\overline{\psi_{\alpha L}} \, h_{\alpha \beta} \, \Phi \, \psi_{\beta R} + 
f_{\alpha \beta} 
\left[ 
\overline{\psi_{\alpha L}^c} \, \epsilon \, 
\vec{\Delta}_L \, \vec{\tau} \, \psi_{\beta L} + (L \rightarrow R) 
\right] \, . 
\ee
Here, $\psi_{\alpha L}$ ($\psi_{\alpha R}$) are the 
left-- (right--)handed lepton 
doublets, $\Phi$ a Higgs bi--doublet and $\Delta_{L, R}$ are 
Higgs triplets. The Yukawa coupling matrices are denoted by $f$ and $h$. 
The presence of two Higgs triplets maintains the 
left--right symmetry and results in a type II see--saw 
mechanism. 
Symmetry breaking is achieved by receiving the following 
vacuum expectation values of the Higgses: 
\be 
\bad 
\langle \Phi \rangle = 
\left( \baz \kappa & 0 \\[0.2cm] 0 & \kappa' \ea \right) & \mbox{ and } & 
\langle \Delta_{L, R} \rangle = 
\left( \baz 0 & 0 \\[0.2cm] 0 & v_{L,R} \ea \right) 
\ea \, . 
\ee
The light and heavy neutrino masses are then obtained by diagonalising 
\be \label{eq:mogem}
\left( \baz m_L           & \tilde{m}_D \\[0.3cm]
            \tilde{m}_D^T & M_R 
 \ea \right)  , 
\ee
where $m_L = f \, v_L$ and $M_R = f \, v_R$ is a 
left--handed (right--handed) Majorana and 
$\tilde{m}_D = h \, \kappa \simeq h \, v$ the Dirac mass matrix, which 
in this scenario is identical to the charged lepton matrix $m_{lep}$. 
The weak scale is $v = 174$ GeV\@. Diagonalization yields 
\be \label{eq:mnu}
m_\nu = m_L - \tilde{m}_D \, M_R^{-1} \, \tilde{m}_D^T \, . 
\ee
Since $v_L \, v_R = \gamma \, v^2$ with $\gamma = {\cal{O}}(1)$, it follows 
that \cite{JPR2} 
\be \label{eq:MRmnu}
M_R = \frac{v_R}{v_L} \, m_L \simeq \frac{v_R}{v_L} \, m_\nu \, , 
\ee
i.e., the low energy mass matrix is identical to the 
high energy matrix, thus the mass spectra are the same at both, 
the see--saw and the low energy scale. 
The matrix $m_\nu$ is further diagonalized by 
\be \label{eq:uldef}
U_L^T \, m_\nu \, U_L = {\rm diag} (m_1,m_2,m_3) \, . 
\ee
$U_L$ is therefore identical to the matrix that diagonalizes $M_R$, 
whose eigenvalues are needed to compute the decay asymmetry. 
This asymmetry is caused by the interference of tree level with one--loop 
corrections for the decays of the heavy Majorana neutrinos, 
$N_i \ra \phi \, l^c$ 
and $N_i \ra \phi^\dagger \, l$: 
\bea \label{eq:eps}
\varepsilon_i = \frac{\D \Gamma (N_i \ra \phi \, l^c) - 
\Gamma (N_i \ra \phi^\dagger \, l)}{\D \Gamma (N_i \ra \phi \, l^c) +  
\Gamma (N_i \ra \Phi^\dagger \, l)} \\[0.5cm]
= \frac{\D 1}{\D 8 \, \pi \, v^2} \frac{\D 1}{\D (m_D^\dagger m_D)_{ii}} 
\D\sum\limits_{j\neq i} {\rm Im} (m_D^\dagger m_D)^2_{ij} \, 
\left( f(M_j^2/M_i^2) + g(M_j^2/M_i^2) \right) \, . 
\eea
The contribution of the heavier neutrinos is washed out and only the 
asymmetry generated by the decay of the lightest one 
(i.e., $M_1$ in normal hierarchies) survives. 
The Dirac mass matrix has been rotated by $U_L$, thus changes to 
$m_D = \tilde{m}_D \, U_L$. The function $f$ stems from vertex 
\cite{vertex} and $g$ from self--energy \cite{self,aporev} contributions: 
\be \label{eq:fapprox}
\baz 
f(x) = \sqrt{x} \left(1 - 
(1 + x) \, \ln \left(\frac{\D 1 + x}{\D x}\right) \right) \; , \, 
& \, g(x) = \frac{\D \sqrt{x}}{\D 1 - x}
\ea \, . 
\ee
For $x \gg 1$ it holds $f(x) + g(x) \simeq -\frac{3}{2\sqrt{x}}$.  
The resonance for degenerate neutrinos with $x \simeq 1$ can be 
corrected to a order unity $\varepsilon_i$ with a resummation 
formalism, which will be commented on below. 
Typically, one expects $v_R$ = ${\cal O}(10^{15})$ GeV and thus 
$v_L$ of order of the light neutrino masses $m_i \ls 0.05$ eV in hierarchical 
schemes.

The decay asymmetry (\ref{eq:eps}) is converted into a baryon asymmetry 
via\footnote{The corresponding formula for supersymmetric scenarios 
is approximately the same, since both, $g^\ast$ and $\varepsilon_1$, are 
enhanced by roughly a factor two.} \cite{leptogenesis} 
\be \label{eq:YB}
Y_B = c \, \frac{\kappa}{g^\ast} \, \varepsilon_1 \, , 
\ee 
with $c = -28/51$, $g^\ast \simeq 110$ and $\kappa$ a dilution factor, 
which can be obtained through solving the Boltzmann equations 
\cite{solv}. There exists a convenient fit \cite{hiki}, which takes 
into account the suppression of $\kappa$ for large $M_1 \gs 10^{14}$ GeV 
and small (large) $\tilde{m}_1 \ls 10^{-5}$ eV 
($\tilde{m}_1 \gs 10^{-2}$ eV), where $\tilde{m}_i$ is defined as 
\be \label{eq:tmi}
\tilde{m}_i = \frac{(m_D^\dagger m_D)_{ii}}{M_i} = 
3.0 \cdot 10^{-5} \, \gamma \, 
\left(\frac{(m_D^\dagger m_D)_{ii}}{\rm GeV^2}\right) 
\left(\frac{10^{-3} \, \rm eV}{m_i}\right) \, 
\left(\frac{10^{15} \, \rm GeV}{v_R}\right)^2 \, \rm eV \, . 
\ee
In addition, the heavy Majorana masses are given by 
\be \label{eq:Mi}
M_i = 3.3 \cdot 10^{13} \, \frac{1}{\gamma} 
\left(\frac{m_i}{10^{-3} \, \rm eV}\right) \,
\left(\frac{v_R}{10^{15} \, \rm GeV}\right)^2 \, \rm GeV \, . 
\ee
The typical numbers one expects are 
$\kappa \simeq 10^{-1} \ldots 10^{-3}$ and 
$|\varepsilon_1| \simeq 10^{-7} \ldots 10^{-5}$, which lead to the 
experimentally observed value \cite{PDG} 
$Y_B \simeq 10^{-11} \ldots 10^{-10}$. 
Since in our scenario $(m_D^\dagger m_D)_{ii}$ is of order 
GeV$^2$, we have for $m_1 = 10^{-3}$ eV and the 
``natural parameters'' $v_R = 10^{15}$ GeV and $\gamma = 1$ the values 
$M_1 \simeq 10^{13}$ GeV and $\tilde{m}_1 \simeq 
10^{-5}$ eV\@. This leads to values of the Yukawa couplings of 
$f \simeq m_\nu/v_L \simeq 0.1 \ldots 1$. We will discuss the case of 
larger and smaller values of $m_i$ later.

\section{\label{sec:estYB}Estimating the baryon asymmetry}
We use the following parametrisation of $U_L$: 
\bea \label{eq:Upara}
U_L = U_{\rm CKM} \; 
{\rm diag}(1, e^{i \alpha}, e^{i (\beta + \delta)}) \\[0.3cm]
= \left( \bad 
c_1 c_3 & s_1 c_3 & s_3 e^{-i \delta} \\[0.2cm] 
-s_1 c_2 - c_1 s_2 s_3 e^{i \delta} 
& c_1 c_2 - s_1 s_2 s_3 e^{i \delta} 
& s_2 c_3 \\[0.2cm] 
s_1 s_2 - c_1 c_2 s_3 e^{i \delta} & 
- c_1 s_2 - s_1 c_2 s_3 e^{i \delta} 
& c_2 c_3\\ 
               \ea   \right) 
 {\rm diag}(1, e^{i \alpha}, e^{i (\beta + \delta)}) \, , 
\eea
where $c_i = \cos\theta_i$ and $s_i = \sin\theta_i$. 
The ``Dirac phase'' $\delta$ appears in terrestrial lepton flavor violating 
processes, whereas the ``Majorana phases'' \cite{Maph} $\alpha$ and $\beta$ 
show up in lepton number violation, e.g., in \onbb{} (\obb). 

The values of $\theta_2$ and \dma{} are known to a good 
precision, corresponding to maximal mixing $\tan^2 \theta_2 
\simeq 0.5 \ldots 2$ and $\dma \simeq (1 \ldots 5) \cdot 10^{-3}$ eV$^2$ 
\cite{BFatm}. Regarding solar neutrinos, 
two solutions are presently favored, the Large Mixing Angle (LMA) 
solution with $\tan^2 \theta_1 \simeq 0.2 \ldots 0.8$ and 
$\dms \simeq (1 \ldots 20) \cdot 10^{-5}$ eV$^2$ and the 
low \dms{} (LOW) solution with $\tan^2 \theta_1 \simeq 0.5 \ldots 1$ 
and $\dms \simeq (0.3 \ldots 5) \cdot 10^{-7}$ eV$^2$ \cite{solBF}. 
Latest data strongly favors the LMA solution \cite{SNO}. 
The third angle $\theta_3$ is bounded to be smaller than about 15 
degrees \cite{petbil}. 
Typical best--fit points are mostly connected with very small 
$\theta_3$ and are $\dma = 2.5 \cdot 10^{-3}$ eV$^2$,  
$\tan^2 \theta_2 = 1$, $\dms = 4.5 \cdot 10^{-5}$ eV$^2$ and 
$\tan^2 \theta_1 = 0.4$ for LMA or $\dms = 1 \cdot 10^{-7}$ eV$^2$ and 
$\tan^2 \theta_1 = 0.7$ for LOW\@.

In the normal mass scheme the eigenstates are ordered as 
$m_3 > m_2 > m_1$ and are given as 
\be \label{eq:minorm}
m_3 = \sqrt{\dma + m_2^2} \; \mbox{ and } \; m_2 = \sqrt{\dms + m_1^2} \, .  
\ee
In a strong hierarchy it holds 
$m_3 \simeq \sqrt{\dma} \gg m_2 \simeq \sqrt{\dms} \gg m_1$. 
In the inverse scheme one has $m_1 > m_2 > m_3$ with 
\be \label{eq:miinv}
m_1 = \sqrt{\dma + m_3^2}  \; \mbox{ and } \; 
m_2 = \sqrt{-\dms + m_1^2}  \, . 
\ee
In case of hierarchical masses, $m_1 \simeq m_2 \simeq \sqrt{\dma} \gg m_3$. 
We will now estimate the leptogenesis parameters with the 
simplifications $\tilde{m}_D = {\rm diag}(0,0,m_\tau)$, 
$\theta_2 = \pi/4$ and keeping only the leading order in $s_3$. 
This has been 
shown to be an excellent approximation within this scenario \cite{JPR2}.

\subsection{\label{sec:norm}Normal hierarchy}
Using the above mentioned simplifications together with 
Eqs.\ (\ref{eq:MRmnu},\ref{eq:eps},\ref{eq:fapprox}) 
we find for the decay asymmetry
\bea \label{eq:epsnorm} 
\varepsilon_1 = - \frac{\D m_\tau^2}{\D 8 \, \pi \, v^2} \, 
\frac{\D 1}{\D t_1 - 2 \, s_3 \, c_\delta}
\left[\frac{\D 1}{\D 1 + t_1^2} \Big( t_1 \, s_{2 \alpha} 
+ 2 \, s_3 (t_1^2 \, s_{2 \alpha + \delta} - s_{2 \alpha - \delta})\Big) 
\frac{\D m_1}{\D \sqrt{\dms}}  \right. \\[0.5cm]
\left. 
+ \left( t_1 \, s_{2(\beta + \delta)} - 2 \, s_3 \, s_{2 \beta}\right) 
\frac{\D m_1}{\D \sqrt{\dma}}
\right] \, , 
\eea
with $s_{2\alpha + \delta} = \sin 2 \alpha + \delta$ and so on. 
Setting $t_1^2 = 1$ one obtains the formulas from \cite{JPR2}. 
The parameter $\tilde{m}_1$ reads 
\be \label{eq:tmnorm}
\tilde{m}_1 = 4.8 \cdot 10^{-5} \, \gamma \, 
\left(\frac{10^{15} \, \rm GeV}{v_R}\right)^2 \, 
\left(\frac{10^{-3} \, \rm eV}{m_1}\right) \, \frac{t_1}{1 + t_1^2} \, 
(t_1 - 2 \, s_3 \, c_\delta) \, \, \rm eV \, . 
\ee
For the best--fit values mentioned in the last section, 
the dilution factor is $\kappa \simeq 0.04$ for $m_1 = 10^{-4}$ eV and 
the  decay asymmetry simplifies to 
\be
-\varepsilon_1 \simeq \left\{ \baz 
4.4 \cdot 10^{-7} \, \left( s_{2 \alpha} + 0.2 \, s_{2(\beta+\delta)} \right) 
\left(\frac{\D m_1}{\D 10^{-3} \, \rm eV}\right)
&  {\rm LMA} \\[0.5cm]
7.7 \cdot 10^{-7} \, s_{2 \alpha} 
\left(\frac{\D m_1}{\D 10^{-4} \, \rm eV}\right) 
&  {\rm LOW} \ea \right. \, . 
\ee
Therefore, for LMA (LOW) values of $m_1$ 
around between $10^{-5}$ to $10^{-3}$ eV 
($10^{-6}$ to $10^{-4}$) eV are required in order to produce 
a sufficient $\varepsilon_1$. 
The smallness of \dms{} for the LOW solution leads to an almost vanishing 
contribution of \dma{} to $\varepsilon_1$ and a larger asymmetry. 
It is seen that for large values of $m_1$ the decay 
asymmetry increases. For $m_1 = 10^{-4}$ eV one finds that 
$Y_B \simeq 10^{-11} (s_{2 \alpha} + 0.2 \, s_{2(\beta+\delta)})$ for LMA 
and $Y_B \simeq 2 \cdot 10^{-10} \, s_{2 \alpha}$ for LOW\@. 
Therefore, values of $\alpha$ around $(2n +1)\pi/4$ are required to produce 
a sufficient asymmetry, which will be confirmed below. 
Values of $m_1$ lower than $10^{-5}$ ($10^{-6}$) eV for 
the LMA (LOW) solution render $\varepsilon_1$ too small and 
represent a rough lower limit on the smallest neutrino mass 
\cite{JPR2}. 
For identical masses the ratio of $Y_B$ for the two solutions reads
\be
\left.\frac{\D Y_B^{\rm LOW}}{\D Y_B^{\rm LMA}}
\right|_{s_3^2 = 0, \; t_1^2 = 1} \simeq 
\frac{\D r_\odot}{1 + s_{2 (\beta + \delta)}/s_{2 \alpha} \, r_{\rm LMA}} 
\simeq r_\odot 
\ , 
\ee
with $r_\odot^2 = (\dms)_{\rm LMA}/(\dms)_{\rm LOW} \gg r_{\rm LMA}^2 = 
(\dms)_{\rm LMA}/\dma \ll 1$. Thus, the baryon asymmetry for the LMA 
solution is smaller by the square root of the 
ratio of the solar mass scales.

\subsection{\label{sec:inv}Inverse hierarchy}
In the inverse hierarchy the lightest neutrino is now $M_3$. There is no 
contribution from the solar scale \dms{} to the decay asymmetry: 
\bea \label{eq:epsinv} 
\varepsilon_3 = - \frac{\D m_\tau^2}{\D 8 \, \pi \, v^2} \, 
\frac{\D m_3}{\D \sqrt{\dma}} \, \frac{\D 1}{\D 1 + t_1^2} 
\left[ s_{2(\alpha - \beta + \delta)} - t_1 \, (t_1 \, s_{2 (\beta + \delta)} 
- 4 \, s_3 \, s_\alpha \, c_{\delta + 2 \beta - \alpha})
\right] \, , 
\eea
We will comment below on the potential enhancement of the decay 
asymmetry due to the degenerate masses $M_1$ and $M_2$. 
For our best--fit values this simplifies to 
\be
-\varepsilon_3 \simeq \left\{ \baz 
5.9 \cdot 10^{-8} \, 
\left(  s_{2(\alpha - \beta + \delta)} - 0.4 \, s_{2 (\beta + \delta)} 
\right) 
\left(\frac{\D m_3}{\D 10^{-3} \, \rm eV}\right)
&  {\rm LMA} \\[0.5cm]
2.1 \cdot 10^{-8} \, 
\left(  s_{2(\alpha - \beta + \delta)} - 0.7 \, s_{2 (\beta + \delta)} 
\right) 
\left(\frac{\D m_3}{\D 10^{-3} \, \rm eV}\right) 
&  {\rm LOW} \ea \right. \, , 
\ee
being one order of magnitude below the values for the normal hierarchy. 
A smaller range for $m_3$ than in the normal hierarchy is found, 
values around $10^{-3}$ to $10^{-2}$ eV are now required: 
we find for $\tilde{m}_3$ that 
\be \label{eq:tminv}
\tilde{m}_3 = \tilde{m}_1 \, \frac{1 + t_1^2}{t_1} \, 
\frac{1}{t_1 - 2 \, s_3 \, c_\delta} \, ,
\ee
which is of the same order of magnitude as $\tilde{m}_1$. Therefore, 
the lower limit on the smallest mass eigenstate, for which 
$\varepsilon_3$ becomes too small, is roughly $10^{-4}$ eV\@. 
For $m_3 = 10^{-3}$ eV, $\kappa$ is about 0.006 and for $t_1^2 = 1$ 
one finds that 
$Y_B \simeq 4 \cdot 10^{-11} \, c_\alpha s_{\alpha - 2(\beta + \delta)}$. 
Therefore, for large $t_1^2$ 
values of $\alpha \simeq n \pi$ are favored, as will be 
confirmed later on. 
If we assume identical smallest mass states and $\kappa$, 
the fraction of the baryon asymmetry in the two hierarchies is 
\be \label{eq:YBnorminv}
\left.\frac{\D Y_B^{\rm norm}}{\D Y_B^{\rm inv}}
\right|_{s_3^2 = 0, \; t_1^2 = 1} \simeq 
\frac{\D s_{2 \alpha} \, r + 2 \, s_{2 (\beta + \delta)}}
{\D 2 \, s_\alpha \,  c_{\delta + 2 \beta - \alpha}} \, , 
\ee
where $r^2 = \dma/\dms \gg 1$. 
Thus, for comparable phases and 
masses, the baryon asymmetry in the normal hierarchical 
scheme is larger by the square root of the  
ratio of the atmospheric and solar mass scales.

\subsection{\label{sec:deg?}Degenerate neutrinos}
The question of leptogenesis with degenerate neutrinos has been 
addressed in the past in different models \cite{deg}. 
As seen from Eqs.\ (\ref{eq:tmi},\ref{eq:Mi}), 
it seems difficult to have degenerate neutrinos 
$m_i \gs 0.1$ eV in our scenario, 
because the resulting large $M_i$ and small $\tilde{m}_i$ 
lead to a strong suppression of $\kappa$. It is however possible to 
decrease $M_i$ by changing $\gamma$ and $v_R$ which would then 
increase $\tilde{m}_i \propto 1/M_i$. As an example, consider $m_i = 1$ eV, 
which for our ``natural parameters'' leads to $M_i \simeq 10^{16}$ GeV 
and $\tilde{m_i} \simeq 10^{-8}$ eV\@. 
If one now decreases $M_i$ by choosing 
$v_R = 10^{13}$ GeV and $\gamma=10$, 
then one has $M_i \simeq 10^{11}$ GeV 
and $\tilde{m_i} \simeq 10^{-4}$ eV, which are again acceptable 
values. The Yukawa couplings turn out to remain basically unchanged, 
$f \simeq m_\nu/v_L \simeq 0.1 \ldots 1$. 
The usual fit of the wash--out parameter $\kappa$ assumes 
hierarchical neutrinos. We will therefore only consider the 
decay asymmetry, which for degenerate neutrinos 
displays a resonance behavior. 
For $M_i^2/M_j^2 \simeq 1$ the vertex part of 
the decay asymmetry is of order $(8 \, \pi \, v^2)^{-1} \simeq 10^{-6}$. 
The precise form of $\varepsilon_i$ for the self--energy 
part reads \cite{aporev}
\be
\varepsilon_i = \sum\limits_{j \neq i} 
\frac{\D{\rm Im} (m_D^\dagger m_D)^2_{ij}}
{\D (m_D^\dagger m_D)_{ii} \, (m_D^\dagger m_D)_{jj}} \, 
\frac{\D \Delta M^2_{ij} \, M_i \, \Gamma_j}
{\D (\Delta M^2_{ij})^2 + M_i^2 \, \Gamma_j^2} \equiv 
\sum\limits_{j \neq i} \phi_{ij} \, \mu_{ij} \, , 
\ee
where we separated the two fractions and introduced the decay width 
\be
\Gamma_i = \frac{\D (m_D^\dagger m_D)_{ii}}{\D 8 \, \pi \, v^2} \, M_i \, . 
\ee 
For $s_3=0$ we find that $\phi_{12} = s_{2 \alpha} \, (1 + t_1^2), \; 
 \phi_{13} = s_{2(\beta + \delta)}$ and 
$\phi_{23} = s_{2(\delta - \alpha - \beta)}$.  
Using (\ref{eq:MRmnu}) and introducing 
$c_i = (m_D^\dagger m_D)_{ii}/(8 \, \pi \, v^2)$, which in our 
scenario is ${\cal O}(10^{-6})$ to an excellent approximation, 
one finds 
\be
\mu_{ij} = 
\frac{\D \Delta m^2_{ij} \, m_i \, m_j \, c_j}
{\D (\Delta m^2_{ij})^2 + m_i^2 \, m_j^2 \, c_j^2} \simeq 
\frac{\D 10^6 \, \frac{\Delta m^2_{ij}}{m_0^2}}
{\D 10^{12} \, \left(\frac{\Delta m^2_{ij}}{m_0^2}\right)^2 
+ 1} \, , 
\ee
where the degenerate mass $m_i \simeq m_j \equiv m_0$ was introduced. 
If $\Delta m^2$ is the atmospheric scale, then 
$\mu_{ij}$ is of order $10^{-6} $ ($10^{-4}$) for $m_0 = 0.1$ (1) eV\@. 
If the scale is $(\dms)_{\rm LMA} \simeq 10^{-5}$ eV$^2$, then 
$\mu_{ij} \simeq 10^{-4} \; (10^{-2})$ for $m_0 = 0.1 \; (1)$ eV\@. 
If $(\dms)_{\rm LOW} \simeq 10^{-7}$ eV$^2$, then 
$\mu_{ij} \simeq 10^{-1}$ independent of $m_0$. 
Therefore, the self--energy part is never larger than order one. 
In the normal hierarchy we find for $m_0=0.1$ eV that 
\be \label{eq:epsdeg1}
10^{6} \, \left|\sum \varepsilon_i \right| \simeq \left\{ \baz 
\left( 10^2 \, s_{2 \alpha} \, (1 + t_1^2) + 
2 \, c_\alpha \, s_{2 (\beta + \delta) - \alpha} \right) 
& \mbox{ LMA}\\[0.4cm] 
 \left( 10^5 \, s_{2 \alpha} \, (1 + t_1^2) + 
2 \, c_\alpha \, s_{2 (\beta + \delta) - \alpha} \right) 
& \mbox{ LOW}\\
\ea \right.
\ee
and for  $m_0=1$ eV 
\be \label{eq:epsdeg2}
10^{6} \, \left|\sum \varepsilon_i \right| \simeq \left\{ \baz 
 \left( 10^4 \, s_{2 \alpha} \, (1 + t_1^2) + 
2 \cdot 10^2 \, c_\alpha \, s_{2 (\beta + \delta) - \alpha} \right) 
& \mbox{ LMA}\\[0.4cm] 
\left( 10^5 \, s_{2 \alpha} \, (1 + t_1^2) + 
2 \cdot 10^2 \, c_\alpha \, s_{2 (\beta + \delta) - \alpha} \right) 
& \mbox{ LOW}\\
\ea \right. \, . 
\ee
Note that $\sum \varepsilon_i$ can not be of order one for the LMA solution. 
If the wash--out factor $\kappa$ is around $10^{-1} \ldots 10^{-3}$ 
as for the hierarchal scheme, the numbers in the right--hand side of the  
last two equations have to be smaller or of the order 1 or 0.1. 
The first term proportional to 
$s_{2 \alpha}$ dominates and can be made small for $\alpha \simeq n \pi/2$, 
which either minimizes or maximizes the second term proportional to 
$c_\alpha$. This demands fine--tuning of $\alpha$ to a 
precision of $10^{-3}$ to $10^{-5}$. The phase has to be closer to $n \pi/2$ 
for the LOW solution and for larger $m_0$. 
The ambiguity in $\alpha$ can be tested in \onbb{} since for 
$t_3^2 \simeq 0$ the effective mass reads 
\be
\meff \!\! \simeq \frac{m_0}{1 + t_1^2} 
\sqrt{(1 + t_1^2)^2 -  4 \; t_1^2 \; s_\alpha^2} 
\simeq m_0 \, c_\alpha\, , 
\ee
where the last approximation corresponds to $t_1^2 \simeq 1$. 
Thus, the cases $\meff \!\! \ll m_0$ and $\meff \!\! \simeq m_0$ 
correspond to 
$\alpha \simeq \pi/2$ and $\alpha \simeq \pi$, respectively. 
Forthcoming experiments \cite{vogel} 
can very well test values of \meff below or around 0.01 eV\@. 

In the last section the potential enhancement of the asymmetry in the 
inverse hierarchy has been mentioned. With the formalism discussed in this 
section, one can calculate now the contribution of the vertex contribution to 
the decay asymmetry from the Majorana neutrinos $M_1$ and $M_2$. 
We find that 
\be \label{eq:epsdeginv} 
\left|\sum \varepsilon_i \right|_{M_1-M_2} \simeq 
10^6 \, s_{2 \alpha} \frac{r^2}{10^{12} + r^4}  \, , 
\ee
with $r^2$ defined after Eq.\ (\ref{eq:YBnorminv}). For LMA, this 
number is of order $10^{-6} \, s_{2 \alpha}$ and for the disfavored LOW 
solution about $10^4$ times this value. One would find again that 
$\alpha \simeq n \pi/2$, which again is testable in \onbb: for 
$t_3^2 m_3 \simeq 0$ and $t_1^2 \simeq 1$ the effective mass reads 
\be \label{eq:meffinv}
\meff \! \! \simeq \sqrt{\dma} \, c_\alpha \; .  
\ee
Thus, the cases $\meff \!\! \ll \sqrt{\dma}$ and 
$\meff \!\! \simeq \sqrt{\dma}$ 
correspond again to 
$\alpha \simeq \pi/2$ and $\alpha \simeq \pi$, respectively. 
Since the masses $M_1$ and $M_2$ are heavier then $M_3$, the asymmetry 
caused by them suffers an additional reduction, we can 
for the inverse hierarchy and the strongly favored LMA solution 
safely work with the conventional form of $\varepsilon_3$ as 
in Eq.\ (\ref{eq:epsinv}).

For very small masses of $m_i \ls 10^{-8}$ eV, which lead to large 
$\tilde{m}_i$ and small $M_i$, it is possible to overcome this problem 
by increasing $v_R$ and decreasing $\gamma$. However, since 
$m_\nu \simeq 10^{-3}$ eV, the resulting values of the Yukawa couplings $f$ 
become now too large and spoil the naturalness of the model. 
We will therefore not discuss this possibility.

\section{\label{sec:res}Numerical Results}
\begin{figure}[ht]
\epsfig{file=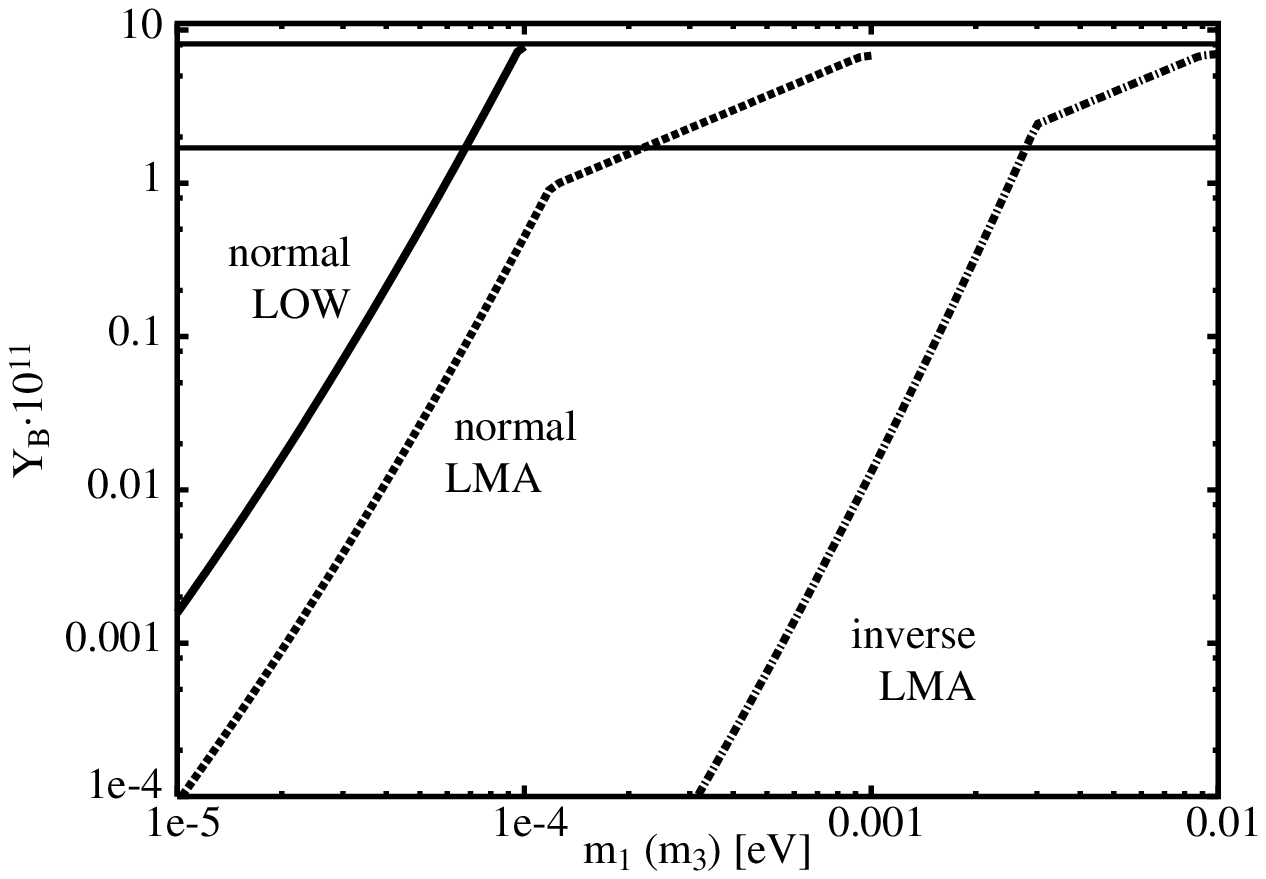,width=13cm,height=8cm}
\caption{\label{fig:YBm}The baryon asymmetry as a function 
of the smallest mass state in both hierarchies for different representative 
values of the Majorana phases, the best--fit points as 
mentioned in the text and $t_3^2 = 0.005$. 
The solid line is for the LOW solution, the dotted for LMA and the 
dashed--dotted for the inverse hierarchy.}
\vspace{1cm}\hspace{-1cm}
\epsfig{file=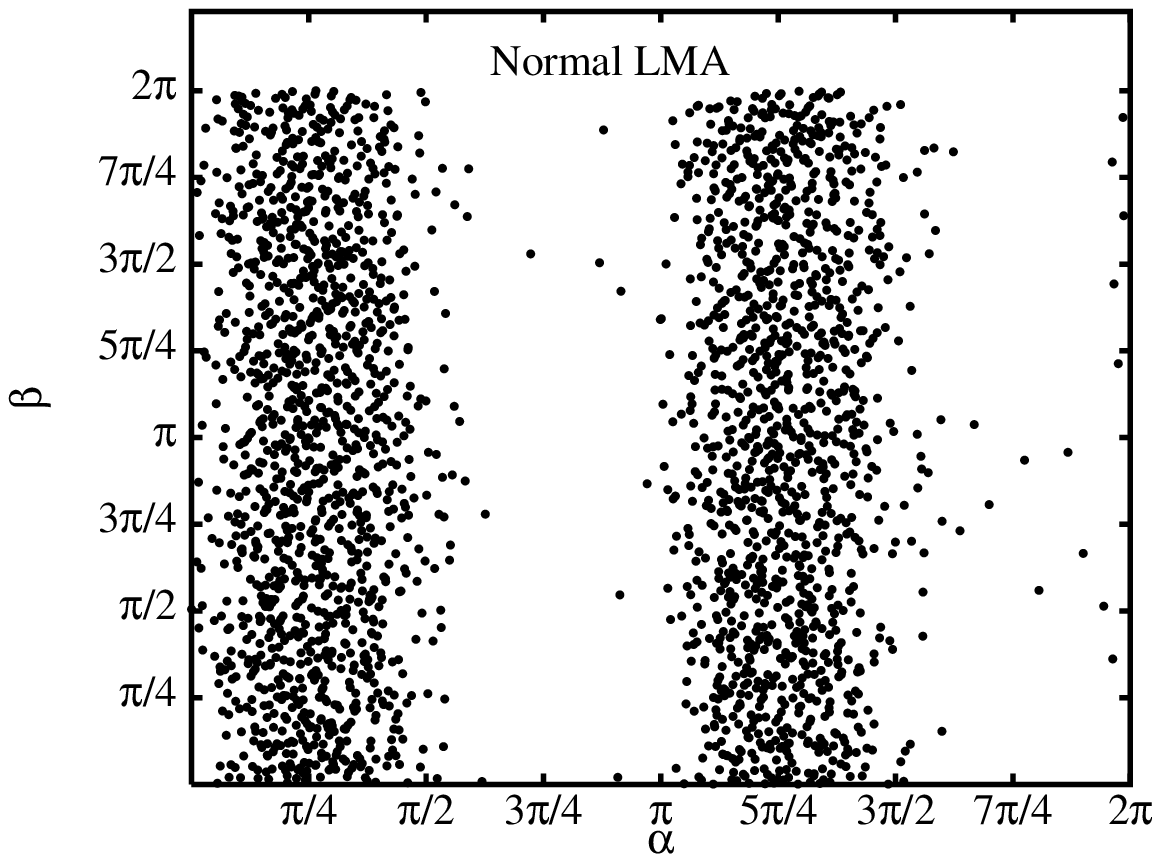,width=5.2cm,height=6cm}
\epsfig{file=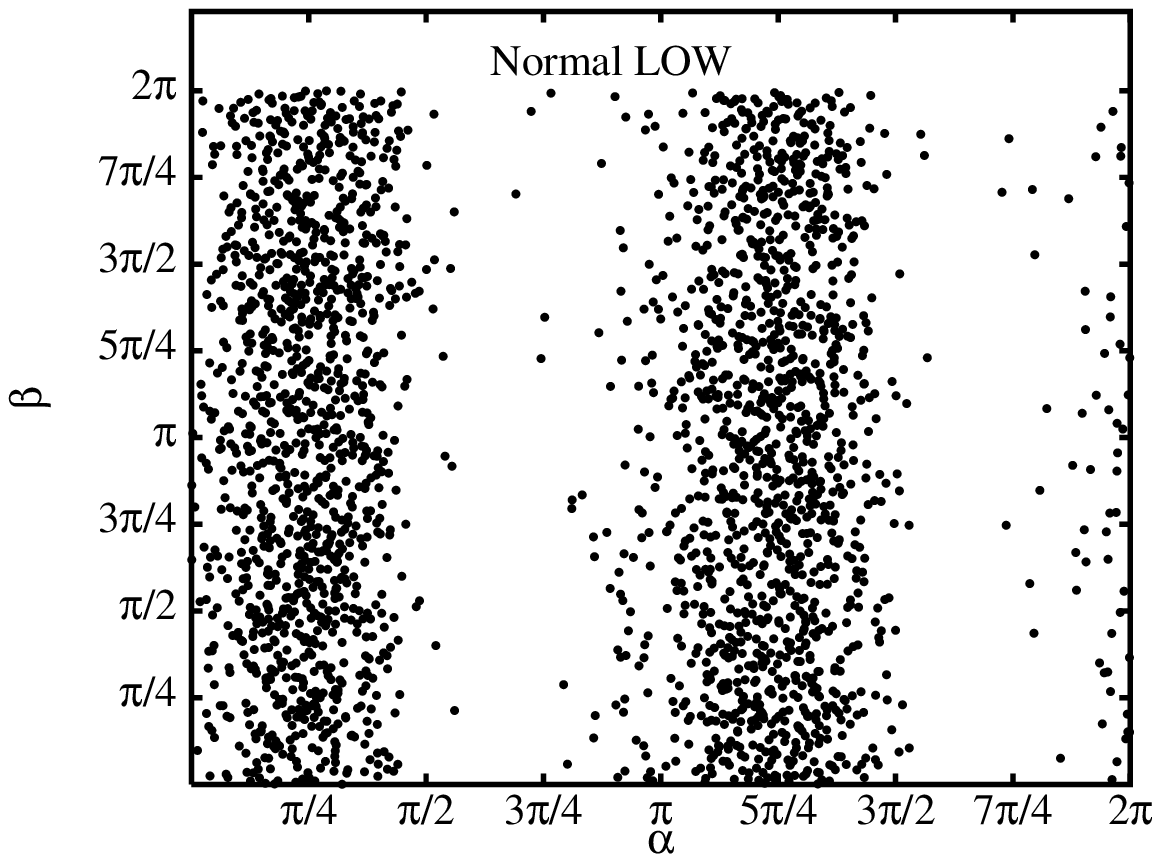,width=5.2cm,height=6cm}
\epsfig{file=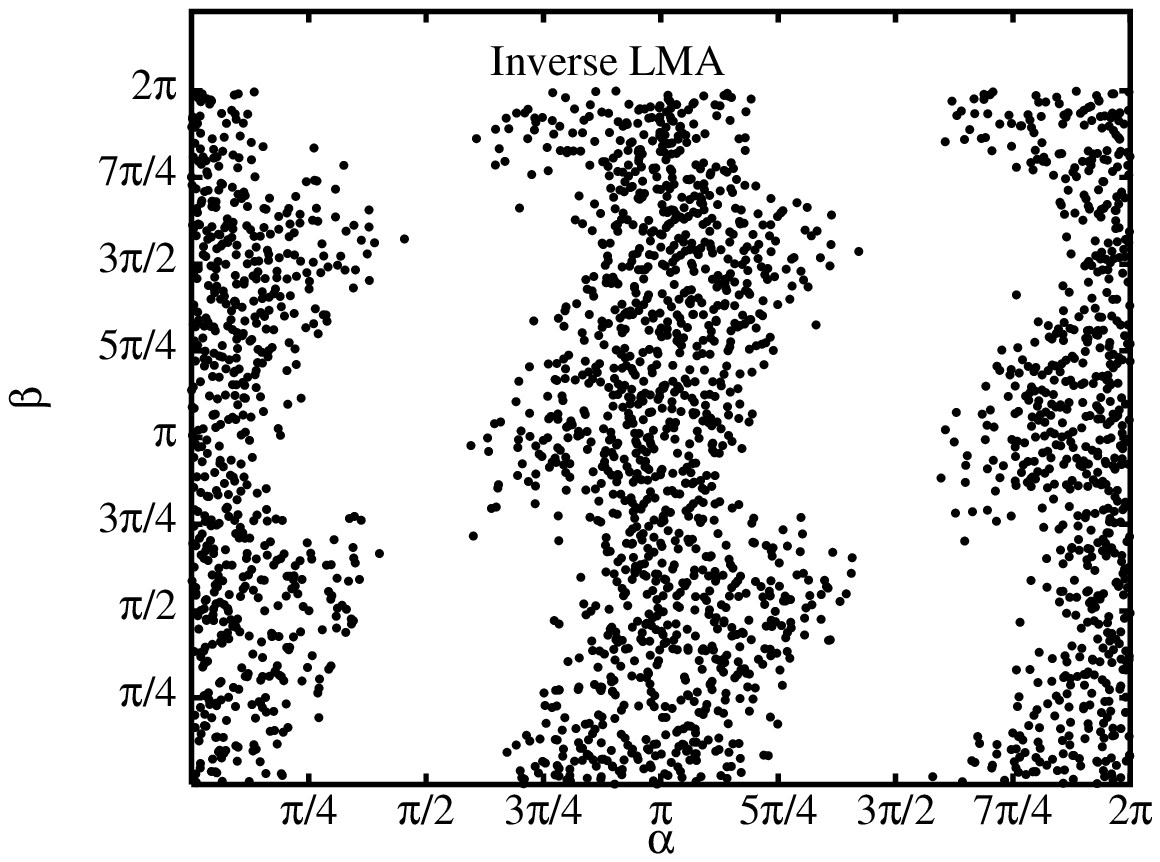,width=5.2cm,height=6cm}
\caption{\label{fig:ab}Scatter plot of the Majorana phases $\alpha$ 
and $\beta$ for both hierarchies.}
\end{figure}

We will now analyze the predictions of our scenario for low energy 
observables. 
For $\tilde{m}_D$ we took 
a typical charged lepton mass matrix 
\be \label{eq:tmdtyp}
\tilde{m}_D = \left( \bad 
0                 & \sqrt{m_e \, m_\mu} & 0 \\[0.3cm]
\sqrt{m_e \, m_\mu} & m_\mu               &  \sqrt{m_\tau \, m_\mu} \\[0.3cm]
0                 & \sqrt{m_\tau \, m_\mu} & m_\tau 
\ea \right) ,   
\ee
where $m_{e,\mu,\tau}$ are the masses of the electron, muon and tau lepton. 
Fig.\ \ref{fig:YBm} shows $Y_B$ as a function of the smallest mass state 
for the best--fit values of the oscillation parameters and $t_3^2 = 0.005$. 
In the inverse hierarchy we took the LMA solution to avoid the 
resonant enhancement as discussed above. 
The preferred value of the smallest mass is larger in the inverse hierarchy 
and larger for the LMA solution. Indicated in the plot are typical 
experimental values of \cite{PDG} 
$Y_B \simeq (1.7 \ldots 8.1) \cdot 10^{-11}$. For the following 
plots we fixed the smallest mass states to 
$10^{-4}$ ($10^{-3}$) eV for 
LOW (LMA and inverse hierarchy).

In Fig.\ \ref{fig:ab} we display scatter plots of the two Majorana phases 
$\alpha$ and $\beta$ for both schemes and solutions. The points are 
obtained by producing random values of the oscillation parameters in the 
range given above and also varying the three phases between zero and $2\pi$. 
When a sufficient baryon asymmetry is produced, the point is kept.  
Due to the 
smallness of $t_3^2$, $\beta$ is basically a free parameter. The 
second phase $\alpha$ lies around $\pi/4$ or $5 \pi/4$ 
in the normal hierarchy\footnote{For fixed values of the 
oscillation parameters the values of the phases can be  
different for the respective solutions, see \cite{apu}. Note however, 
that the dependence on $m_1$ of $\kappa$ has not been fully 
taken into account in that analysis.} 
and around $0$, $\pi$ or $2\pi$ in the inverse hierarchy. These 
values confirm our estimates in Section \ref{sec:estYB}. 
In the normal hierarchy, \meff is a function of $(\alpha - \beta)$, 
to be precise: 
\be
\meff \!\!^2 \simeq 
\dms s_1^4 + \dma t_3^4 +2 \, s_1^2 \, t_3^2 \, \sqrt{\dms \dma} \, 
c_{2(\alpha - \beta)}~. 
\ee
For sizable \meff and a sizable 
dependence of \meff on the phases large 
values of $t_3^2$ are required. 
It turns out that 
$\alpha = \pi/4$ or $5 \pi/4$ yield similar results for \meff 
and that for large 
$t_3^2$ values of $\beta \simeq 3 \pi/4$ lead to unobservable \meff$\!\!$. 
See \cite{apu} for details.

\begin{figure}[ht]
\epsfig{file=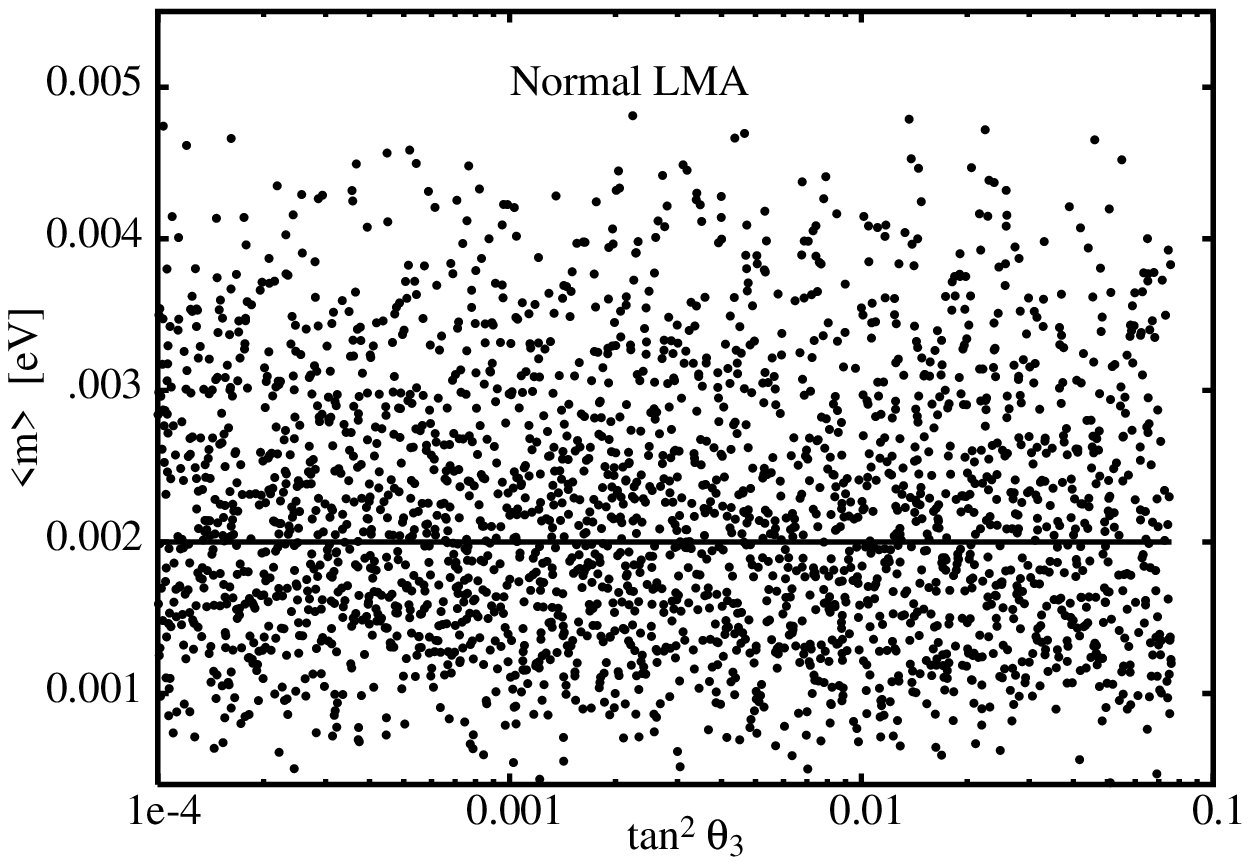,width=7.5cm,height=7cm}
\epsfig{file=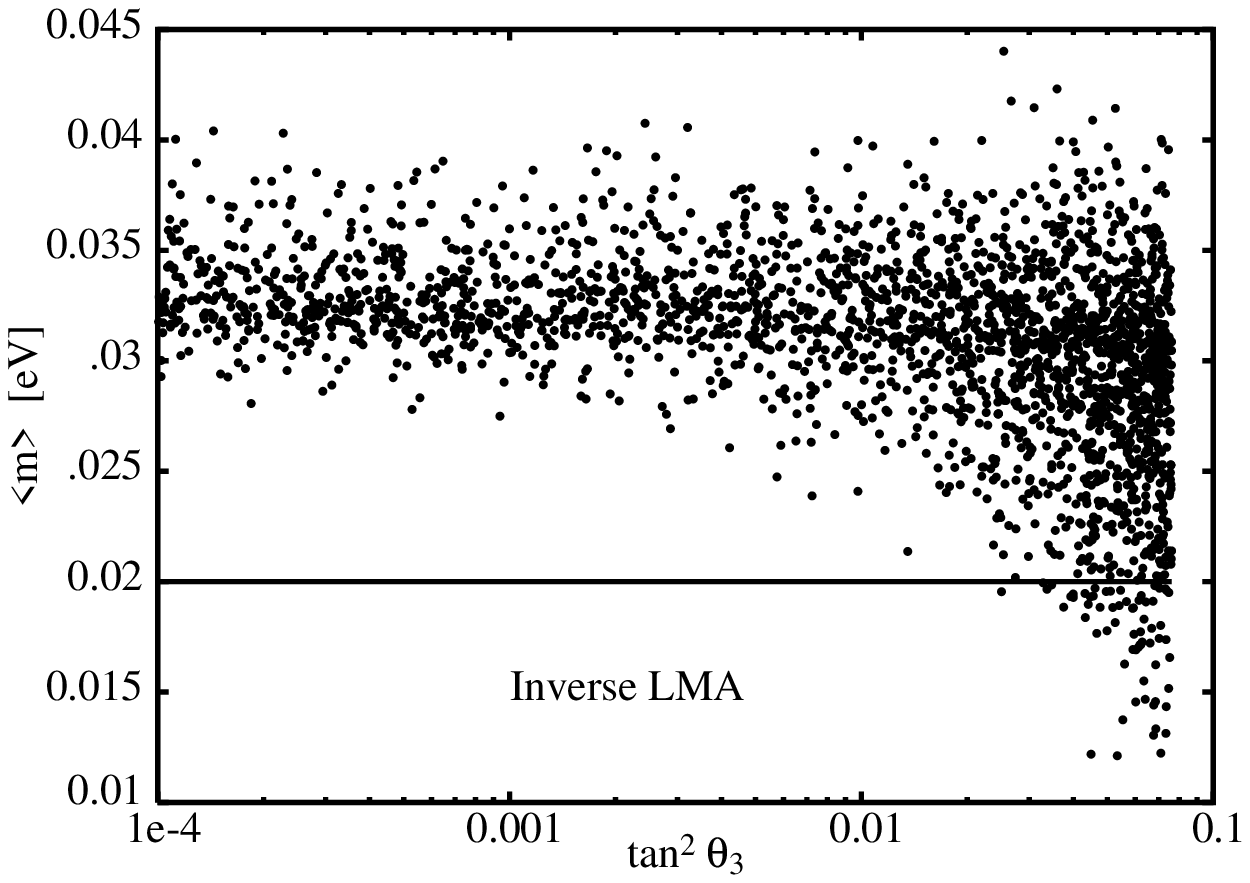,width=7.5cm,height=7cm}
\caption{\label{fig:meff}Scatter plot of $\tan^2 \theta_3$ 
and \meff for both hierarchies in the LMA solution.}
\vspace{2cm}
\epsfig{file=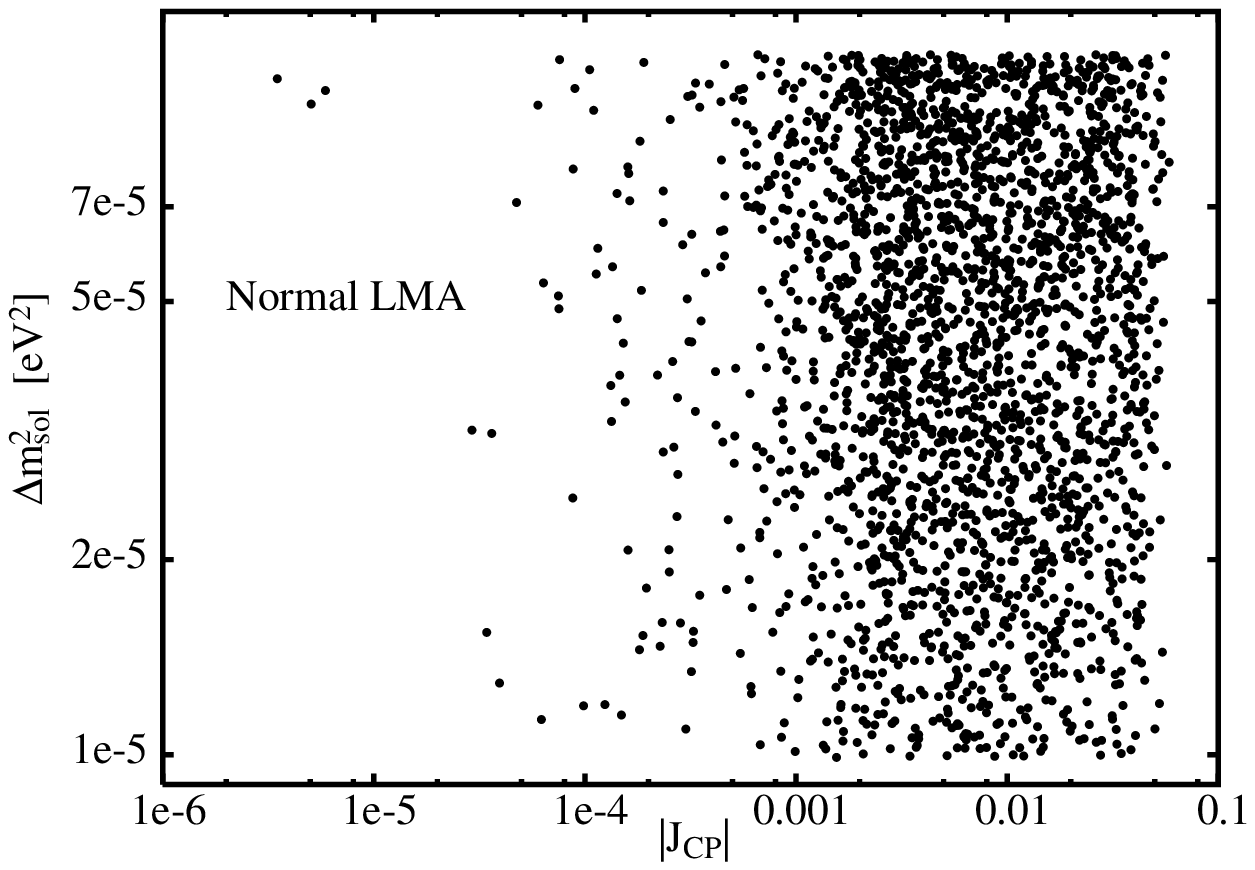,width=7.5cm,height=7cm}
\epsfig{file=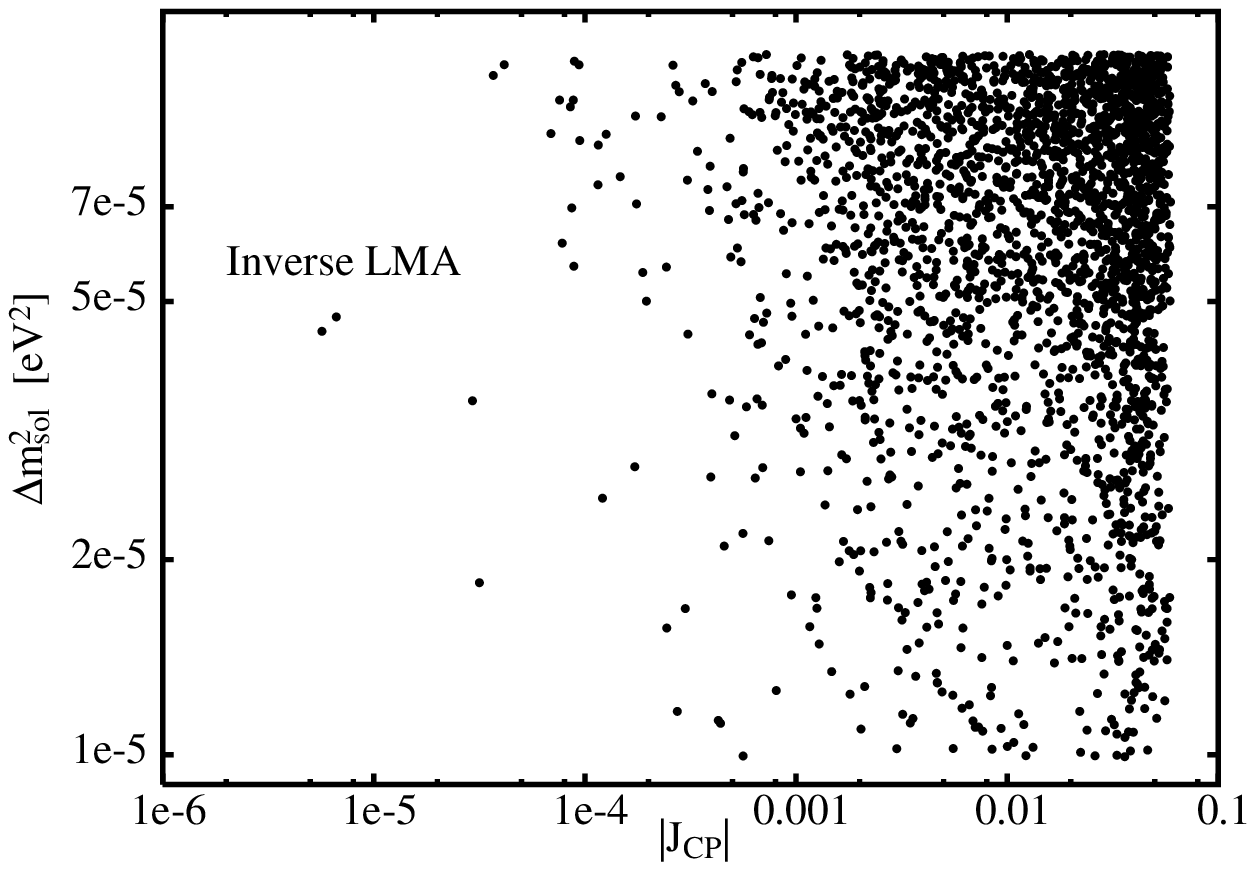,width=7.5cm,height=7cm}
\caption{\label{fig:JCP}Scatter plot of $J_{CP}$ 
and \dms{} for both hierarchies in the LMA solution.}
\end{figure}

In the inverse hierarchy however, the preferred 
values of $\alpha$ are of particular interest for \obb. 
From Eq.\ 
(\ref{eq:meffinv}) it is seen that for values of $\alpha = n \pi$ 
there are no cancellations in \meff and 
it holds $\meff \!\! \simeq \sqrt{\dma}$. This is shown  
in Fig.\ \ref{fig:meff}, where scatter plots of $t_3^2$ and \meff are shown. 
Practically all points for the inverse hierarchy lie above 0.02 eV, a value 
observable by future experiments \cite{vogel}. In the normal 
hierarchy, approximately half of the points lie above 0.002 eV, which is 
a very ambitious limit planned to be achieved by the GENIUS experiment 
\cite{GENIUS}. From the figure it becomes also clear, that 
the inverse hierarchy prefers large values of $t_3^2$, since most 
of the points lie above 0.01.

Finally, in Fig.\ \ref{fig:JCP} we display \dms{} against the $CP$ violating 
leptonic Jarlskog invariant $J_{CP}$, which is defined as 
\be 
J_{CP} = \frac{1}{8} \, \sin 2 \theta_1 \, \sin 2 \theta_2 
\, \sin 2 \theta_3 \, \cos \theta_3 \, \sin \delta 
\ee
and may be measured in next generation neutrino experiments \cite{JCP}. 
Necessary conditions are that LMA is the solution of the solar 
neutrino problem and that \dms{} is not too small. 
It is seen from the plots that there is a slight preference for 
large \dms{} in the normal hierarchy and a very strong one in the 
inverse hierarchy. Also, due to the large $t_3^2$ values, $J_{CP}$ is 
larger in the inverse hierarchy.

It is an interesting question to ask how the baryon asymmetry 
depends on the two different kinds of phases, the Dirac phase 
$\delta$ and the Majorana phases $\alpha$ and $\beta$. In 
the minimal $SO(10)$ model analyzed in \cite{branco2} it has been 
observed that a sufficient baryon asymmetry can be produced alone 
with one single Majorana phase, whereas the Dirac 
phase alone is not enough. Reference \cite{ellis}, analyzing the 
minimal supersymmetric see--saw model, 
observes that leptogenesis is insensitive to the Dirac phase. 
From Eq.\ (\ref{eq:epsnorm}) one notes that in the normal hierarchy 
and the LOW solution, there is hardly any dependence on $\beta$ and 
$\delta$ and for the LMA solution the dependence is suppressed. 
The situation is thus similar to the one in \cite{branco2}. 
In the inverse scheme and the LMA solution however, 
the phases can contribute comparably to $Y_B$, 
as can be seen from Eq.\ (\ref{eq:epsinv}). 
Therefore, the Dirac phase alone is sufficient to produce the 
required amount of baryon asymmetry. In case of the LOW solution and also 
for the degenerate scheme, the Majorana phase $\alpha$ is dominant 
and the other phases play no role. It would be interesting to 
investigate this behavior in other models. Naively one expects minor 
dependence of the baryon asymmetry on $\delta$, since it appears in most 
parameterizations with the small quantity $s_3$. 
In addition, since leptogenesis requires lepton number violation, the 
Majorana phases can be expected to play the major role.

A final comment concerns the gravitino problem of leptogenesis, 
which appears once one 
embeds a theory in a supergravity framework. Majorana neutrinos 
with masses considerably below $10^{10}$ GeV can potentially 
evade this problem. 
The resulting smallness of $\kappa$ is fixed by the resonant 
behavior of $\varepsilon_i$. For instance, $M_i \simeq 10^8$ $(10^4)$ GeV 
together with $m_i \simeq 1$ eV leads to $v_R \simeq 10^{11}$ $(10^9)$ GeV, 
$v_L \simeq 10^3$ $(10^5)$ eV and $f \simeq 10^{-4}$ $(10^{-6})$, when 
$m_\nu \simeq 0.1$ eV is assumed. The mass 
$\tilde{m}_i$ will then be of the order of 1 ($10^4$) eV\@. 
Extrapolating $\kappa$ for 
Majorana masses of order $10^8$ GeV leads to 
$\kappa \simeq 10^{-6}$, thus 
$\sum \varepsilon_i \simeq 10^{-3} \ldots 10^{-2}$. 
Therefore, extreme fine--tuning of $\alpha \simeq \pi/2$ 
is required, leading to very small \meff$\!\!$. 
Lower values of $M_i$ as in TeV scale leptogenesis \cite{TeV} 
with high values of $\tilde m_i$ 
lead to extremely small values of $\kappa$ and 
demand $\sum \varepsilon_i$ to be of order one, in conflict 
with the favored LMA solution.

\section{\label{sec:concl}Conclusions}
Leptogenesis in left--right symmetric models is investigated for all 
possible neutrino mass schemes. Depending on the 
solar solution and the mass scheme, the preferred value of the 
lowest mass state differs. In addition, predictions of low energy 
observables, such as \meff and $J_{CP}$ differ. Especially in the 
inverse hierarchy large values of $t_3^2$ are required, which 
predict observable $CP$ violation in oscillation experiments. 
Furthermore, only little cancellation in \onbb{} is predicted, 
which leads to $\meff\!\! \simeq \sqrt{\dma}$. 
The degenerate scheme predicts $\alpha$ to be around $\pi/2$ 
or $\pi$, which can easily be tested in next generation 
\obb{} experiments. The condition $\alpha \simeq \pi/2$ 
holds for the case of 
Majorana neutrinos with masses not much below $10^{10}$ GeV, as 
required in order to evade the gravitino problem.

\hspace{3cm}
\begin{center}
{\bf \large Acknowledgments}
\end{center}
I thank Anjan Joshipura for helpful comments and careful reading of the 
manuscript. This work has been supported by the
``Bundesministerium f\"ur Bildung, Wissenschaft, Forschung und
Technologie'', Bonn under contract No.\ 05HT1PEA9.

\end{document}